\documentclass[nologo,11pt,a4paper,hidelinks]{ETHpaper}

\usepackage{booktabs}
    
\usepackage{siunitx}

\usepackage{graphicx}

\usepackage{caption}
\usepackage[numbers]{natbib}     %

\usepackage{enumitem}
\setlist{nosep}
\usepackage{amsmath}

\usepackage[capitalise,noabbrev]{cleveref}
\usepackage[]{csquotes}
\renewcommand{\phi}{\varphi}

\title{Resilience: \\ Understand Breakdown, Foster Recovery, and \\ Choose the Right Perspective\footnote{Keynote at the 2026 Political Networks Conference,
University of Manchester, UK, 4-7 August 2026  (Extended version. For a literature overview see the cited papers, in particular \citep{Schweitzer-Andres-ea-2022-modelingsocialresilience}).}}
\titlealternative{Resilience: Understand Breakdown, Foster Recovery, and \\ Choose the Right Perspective}
\author{Frank Schweitzer}
\authoralternative{Frank Schweitzer}
\address{ETH Zurich, Switzerland\\
  \url{www.sg.ethz.ch}}
\www{\url{http://www.sg.ethz.ch}}

\makeframing

\begin{document}

\renewcommand{\thefootnote}{\fnsymbol{footnote}}

\maketitle

\begin{abstract}
  Resilience denotes the capacity of a system to withstand shocks and to recover from them. We distinguish between two different types of dynamics. The first allows for a separation between phases of normalcy and phases of rapid breakdown followed by slow recovery. The second applies to volatile organizations in which such phases are intertwined.

Breakdown is often self-inflicted. Situation awareness is impaired by psychological mechanisms that lead to incorrect expectations regarding societal dynamics. Through positive feedback, the failure of a few elements is amplified into a failure cascade. However, positive feedback can also be harnessed to enable recovery.

In volatile systems, resilience must be understood as an emergent property arising from the interaction of agents. This necessitates a data-driven approach to inform agent-based models, drawing on repositories, knowledge graphs, or tools from artificial intelligence. Such models help demonstrate that resilience represents a compromise between robustness and adaptivity. Maximizing performance frequently comes at the expense of resilience, and second-order solutions aimed at transforming the system prove more promising than attempts to reconstruct past conditions.
\end{abstract}

\section{The Magic Word: Resilience}
\label{sec:introduction}

A sudden change is rarely welcome.
In England, there is the famous nursery rhyme about
{Humpty Dumpty}.
The precarious egg \enquote{sat on the wall} and \enquote{had a great fall}.
The tragedy of the outcome: nobody was able to \enquote{put Humpty together again.}
{Humpty Dumpty} was certainly not resilient.
Resilience requires two properties: the capacity to \emph{withstand} shocks \emph{and} the ability to \emph{recover} from them.
Humpty Dumpty failed on both counts.
This stands in stark contrast to what Nelson Mandela once said:
\enquote{Do not judge me by my success, judge me by how many times I fell down and got back up again.}
That is resilience in essence.

In this respect, resilience differs from stability, which only considers robustness against shocks, whereas resilience encompasses the process of recovery that follows \emph{after} a shock.
Resilience also differs from performance, as peak performance frequently comes at the cost of \emph{reduced} resilience, as burnout, for example, illustrates.
Finally, resilience should not be conflated with flexibility.
{FOMO}, the fear of missing out, often drives individuals and organizations to respond rapidly to all manner of trends, thereby losing their focus and diluting their core business.

If one speaks more broadly about change rather than shocks, resilience comprises two independent dimensions: robustness and adaptivity, the ability to respond to change adequately. 
The question that must be asked is: are our social and economic systems resilient? \cite[]{Carpenter2001,folke2006}.
This leads to a deeper question, how should these systems be described in the first place?

\begin{figure}[t]
  \centering
  \includegraphics[width=0.45\textwidth]{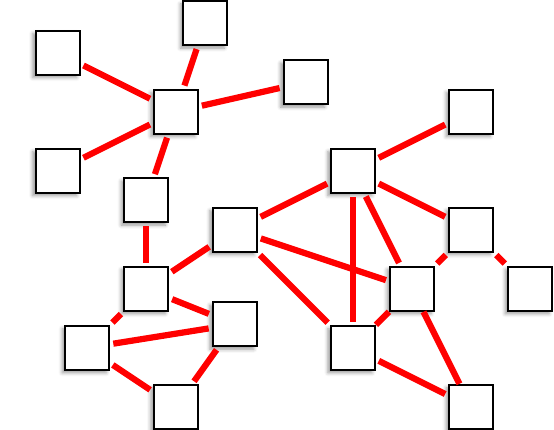}
  \caption{System representation as a network. Nodes denote system elements and links their connections. Nodes may encapsulate entire subsystems.}
  \label{fig:systems}
\end{figure}

\section{Systems}
\label{sec:system}

There is broad consensus on the necessity of adopting a systemic approach. However, what this means, both conceptually and with respect to modeling, remains a matter of considerable debate \citep{Schweitzer-2022-agentsnetworksevolution}. 
A system must be delineated as an entity distinct from its surrounding environment, but it must also be defined in terms of its constituent elements. In most cases, these distinctions give rise to a hierarchy of systems of systems.
This raises a fundamental question: does treating problems at such an abstract level yield genuine insight, or does the absence of concrete specification cause more to be lost than gained?

Systems share essential structural features with respect to elements and their relations, as sketched in \cref{fig:systems}. 
Systems are comprised of system elements, and \emph{complex systems}, in particular, consist of a large number of strongly interacting elements, often referred to as \emph{agents}. The complex systems perspective seeks to explain how systemic properties \emph{emerge} from the interactions among agents, that is, how collective properties arise that cannot be attributed to individual agents alone. This idea is commonly expressed by the aphorism attributed to {Aristotle}: \enquote{The whole is greater than the sum of its parts.}

In this sense, resilience is a \emph{systemic property}. Modeling it within the complex systems framework poses a number of challenges. Sufficient information is required regarding the agents, their individual properties, and their potential interactions. Notably, \emph{complex} agents are not a prerequisite for complex collective dynamics. Nevertheless, it is necessary to characterize their \emph{heterogeneity}. That is, the degree of variation in individual properties, as well as the \emph{network} structure governing their interactions. Such information is often difficult to obtain. Agent-based modeling therefore relies on assumptions about interaction rules, whereas data-driven modeling attempts to infer such rules from available data.

In the absence of data at the \enquote{micro} level of individual agents, the dynamics of the system can still be described at the \enquote{macro} level. This allows a different set of questions to be addressed, concerning, for example, feedback cycles, critical transitions such as regime shifts, and the influence of certain global parameters.

\begin{figure}[t]
  \centering
\includegraphics[width=0.35\textwidth]{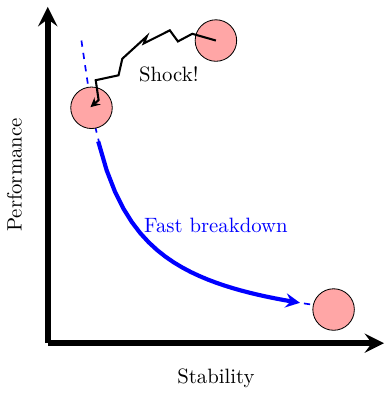}(a)     
\hfill
\includegraphics[width=0.35\textwidth]{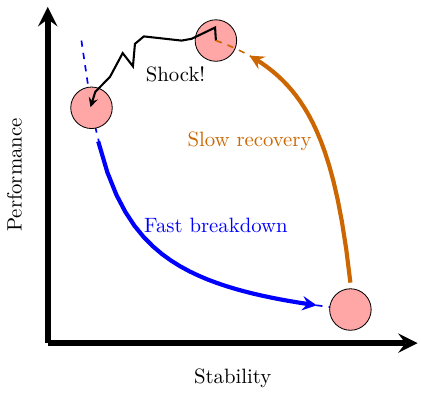}(b)
  \caption{(a) The tragedy of the social planner: Well-performing systems are unstable and prone to breakdown, whereas poorly performing systems tend to exhibit greater stability. (b) While breakdown is fast, recovery is slow and requires resources.}
  \label{fig:recover}
\end{figure}

\section{Tragedy of the Social Planner}
\label{sec:planner}

Researchers across disciplines maintain their own formal or informal descriptions of the systems they investigate, along with detailed accounts of the problems those systems face: radicalization, polarization, inequality, susceptibility to misinformation, and so forth.

From the perspective of systems design \citep{Schweitzer-2020-designing}, such systems can be examined at a rather abstract level. Consider a social planner, a government or central authority, tasked with designing or influencing a system such that it maximizes some abstract performance criterion. In economics, this criterion might be GDP per capita. In social systems, the social planner may seek to maximize cooperation within society in order to improve social welfare, to strengthen democratic institutions so as to enable the participation of all citizens, or to maximize equal access to the commons.

These are valuable and legitimate goals. Yet the social planner ultimately fails. While improvements to the system state, reductions in inequality, increases in the level of cooperation, and similar gains, may well be achieved, the system subsequently encounters what will here be termed a shock: an administrative scandal, a sudden oil price jump, or a comparable disruption. As a result, the carefully designed system state loses its stability and collapses rapidly. In place of the well-functioning system, one finds a failed state, an autocratic rather than a democratic government, a corrupt administration, or a citizenry that has lost trust in its elected representatives. From an abstract performance perspective, these resulting system states are all inferior, they are capable of performing considerably better. Nevertheless, they tend to be relatively stable.

The system that is desirable performs well but proves unstable, whereas the system that actually persists performs poorly but resists change. This constitutes a highly abstract characterization of a recurrent pattern observed across real-world systems.
The central questions are therefore: why does this pattern arise, and what, if anything, can be done to address it?

\section{Systemic Risk: The Big Collapse}
\label{sec:systemic-risk}

Before discussing the breakdown itself, we should first consider the nature of the shock.

If a shock is capable of affecting an entire system, it is referred to as a systemic risk. Classic examples of such shocks are natural catastrophes, such as floods or earthquakes, known as \emph{extreme events} \citep{Albeverio-Jentsch-Kantz-2006-extremeevents,Tessone-Garas-ea-2012-how}. Both the insurance industry and public administration have an interest in estimating, for instance, the probability of an earthquake, more precisely, the \emph{cumulative probability distribution} for earthquakes exceeding a threshold magnitude, in a given region \cite[]{chavez-demoulin-embrechts-hofert-2015-extrem-value}.
Depending on local conditions, such as the robustness of infrastructure or buildings, one can then estimate what fraction of the \enquote{system}, i.e. the municipality, is likely to be destroyed. This does not provide details about \emph{when} and \emph{where} such catastrophes may occur, but it nonetheless offers a useful indication of what one should be prepared for.

\begin{figure}[t]
  \centering
\raisebox{20pt}{\includegraphics[width=0.33\textwidth]{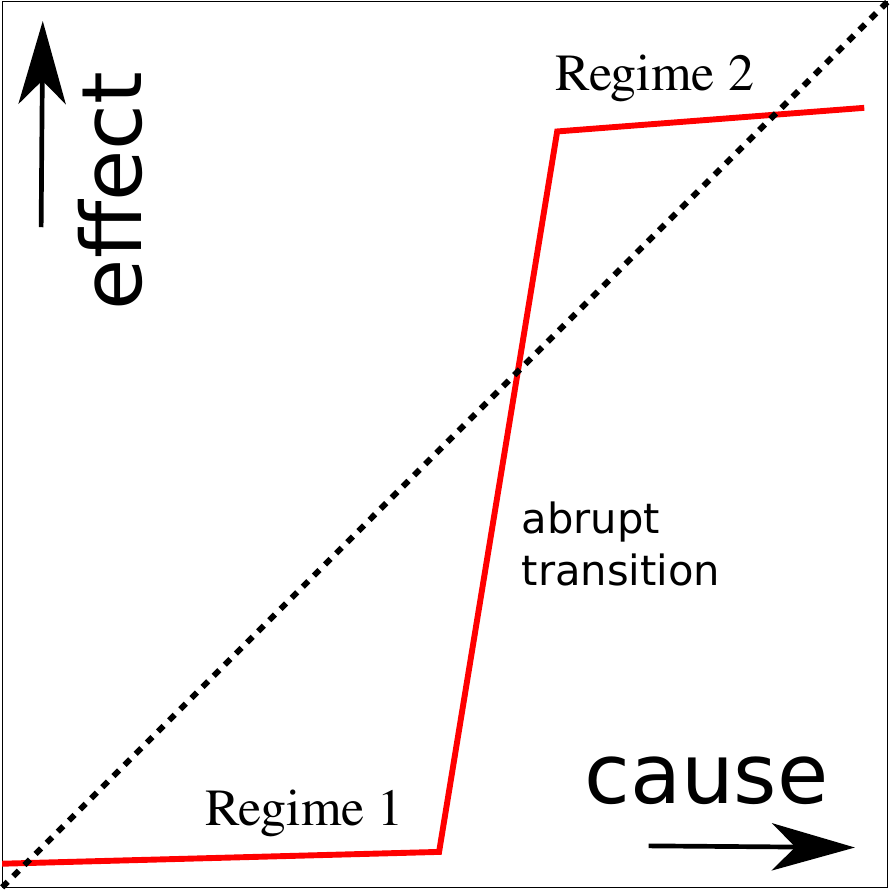}}(a)
  \caption{(a) Nonlinear relation between cause (shock) and effect (breakdown). Below a critical threshold of the shock the system is relatively stable (Regime 1), above this critical threshold an abrupt transition toward Regime 2 occurs.}
  \label{fig:shift}
\end{figure}

Note that in this case, risk is treated as \emph{exogenous} to the system. The earthquake impacts the system, but the collapsing system has no feedback on the probability of such events, at least within certain limits. It is therefore appropriate to describe the system from a \enquote{macro} perspective. This stands in sharp contrast to those cases where systems generate the conditions for their own failure. For instance, when a few elements in a power grid fail, this can cause further elements to fail in turn. A failure cascade may then develop, as the failure of \enquote{the few} becomes amplified through internal or external feedback processes. Epidemic spreading, as well as the insolvency of institutions during the financial crisis, belong to this category. Here, systemic risk is \emph{endogenous} to the system. Understanding the \emph{emergence} of such failure cascades clearly requires models that capture the \enquote{micro} level \citep{Lorenz-Battiston-Schweitzer-2009-systemic,Burkholz-Garas-Schweitzer-2016-how,Burkholz-Leduc-ea-2016-systemicrisk}. Ultimately, the goal is always to quantify systemic risk, that is, to determine what fraction of a system is affected or will survive, but the models required will differ substantially between the two cases.

In both cases, whether shocks are exogenous or endogenous, a nonlinear relationship between cause and effect is observed (\cref{fig:shift}a). Below a critical threshold, the system remains in a relatively stable state, \emph{Regime~1}, and no transition occurs. If the shock exceeds this threshold, however, a sharp transition is observed, an abrupt change to \emph{Regime~2}. This transition is not continuous; rather, the system can be found in distinct phases, each persisting for a characteristic duration.

\section{A Tale of Time Scales}
\label{sec:systems-time-scales}

Investigating systemic risks is only a first step toward understanding \emph{resilience}.
The ability to withstand shocks was previously referred to as the \emph{robustness} dimension of resilience.
Regardless of whether these shocks are large or small, systems that lack the ability to \emph{recover} from them, that is, to regain their previous level of functionality, are \emph{not resilient}.
The second dimension of resilience, \emph{adaptivity}, is therefore of equal importance.
It characterizes how readily a system can recover following a breakdown.

\begin{figure}[t]
  \centering
      \includegraphics[width=0.45\textwidth]{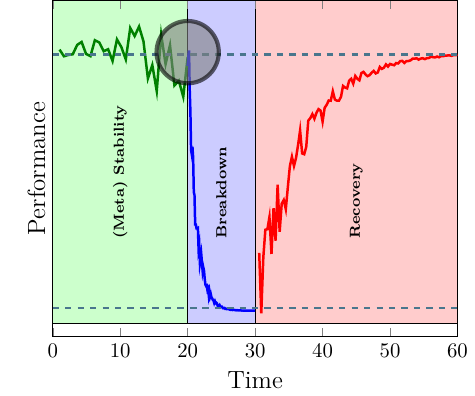}(a)
\hfill
\includegraphics[width=0.45\textwidth]{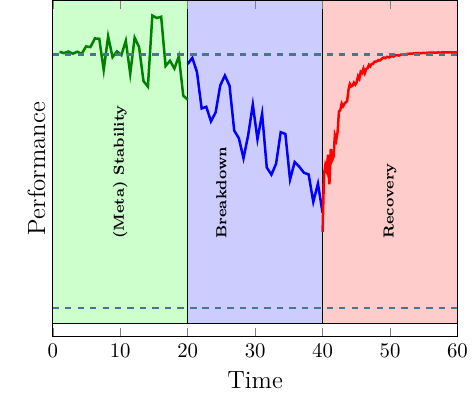}(b)
  \caption{(a)Identification of three phases: metastability, breakdown and recovery. (b) Aim of intervention policies: Extend the time scale of breakdown, to prevent the worst case scenario and to allow for faster recovery.}
  \label{fig:scales}
\end{figure}

Addressing such questions requires a more precise definition of the terms \enquote{breakdown} and \enquote{recovery}, and, by extension, of what constitutes the normal state of a system.
\cref{fig:scales}(a) serves to illustrate this problem.
The figure shows the dynamics of a system over time, in which three distinct phases can be identified.
The green phase represents a \enquote{normal} period of relative stability, during which the system performs at a reasonably high level with moderate fluctuations.
It is referred to as metastable, since it does not represent an absolutely stable state.
The black dot marks the onset of a shock, which is followed by a rapid breakdown during the blue phase.
As this breakdown is not interrupted, the system deteriorates to its lowest level of functionality.
After a prolonged period, the system begins to recover, albeit slowly, as indicated by the red phase, and only gradually returns to its previous level of functionality.

These \emph{three distinct phases}
constitute the classical framework for resilience studies in engineering.
A bridge remains relatively stable over time, even as gradual erosion progresses undetected.
A subsequent earthquake destroys the bridge, with the pre-existing erosion potentially contributing to the collapse.
The decision to rebuild the bridge, sometimes with increased capacity, is then made by external authorities.
In this setting, both the cause of the collapse and the effort required for recovery are external to the \enquote{system}, i.e., the bridge itself.

The clear separation of time scales serves to define the objective of intervention policies.
As \cref{fig:scales}(b) illustrates, the goal is to extend the duration of the blue phase, thereby prolonging the time scale of the breakdown and creating a window for intervention.
If successful, the system would not deteriorate to its lowest level of functionality, enabling recovery to commence from a less severe state and potentially proceed more rapidly.
This picture closely corresponds to the 2008 financial crisis, during which central banks were effectively purchasing \emph{time} in order to slow the pace of the breakdown.
Not all institutions could be preserved in this process; some smaller banks were allowed to fail, a loss that was deemed acceptable as the price of buying time.

\begin{figure}[t]\centering
  
  \includegraphics[width=0.29\textwidth]{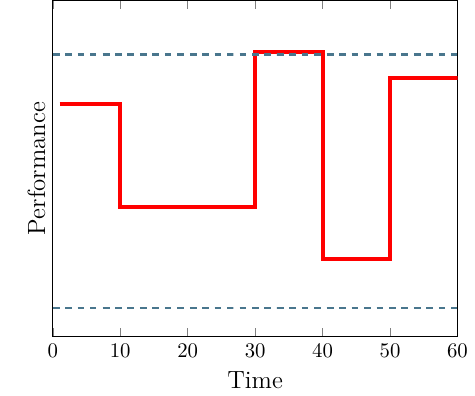}(a) \hfill
  \includegraphics[width=0.29\textwidth]{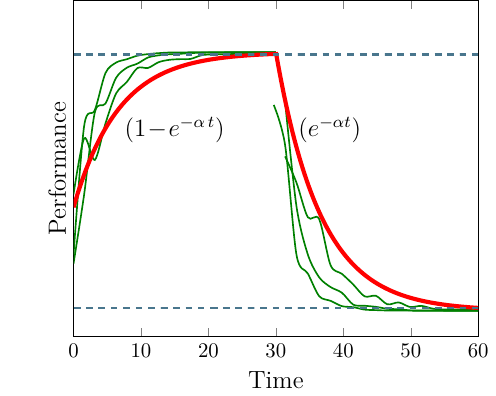}(b) \hfill
  \includegraphics[width=0.29\textwidth]{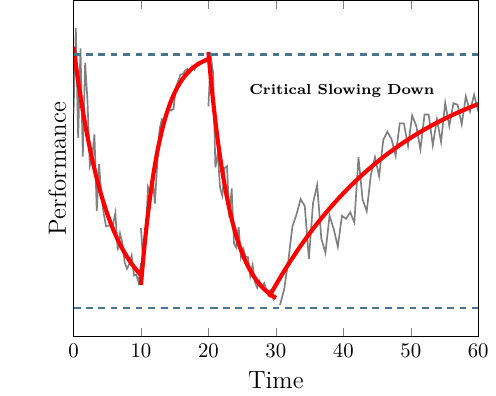}(c) 

  \caption{Three limiting cases of the dynamics shown in \cref{fig:scales}. (a) Comparative statics, in which changes are assumed to occur infinitely fast, whereas equilibrium states persist over long periods. (b) The slow process (red) governs the overall dynamics, while the fast processes (green) relax into a quasi-stationary equilibrium controlled by the slow process. (c) Critical slowing-down in the recovery dynamics impedes the return to a normal state. }
  \label{fig:limit}
\end{figure}

\cref{fig:limit} illustrates three limiting cases of this dynamics, which have been extensively studied across different scientific disciplines.
\cref{fig:limit}(a) shows an example of comparative statics, as employed in economics. A control parameter, such as price, is varied, causing the economic system to transition to a different equilibrium between supply and demand. The underlying assumption is that this market adjustment occurs sufficiently rapidly that supply and demand can be treated as responding instantaneously, rendering the adjustment process itself irrelevant, only the shift between equilibria is of interest.

\cref{fig:limit}(b) illustrates a situation in which multiple concurrent dynamic processes compete for control over the system. Some of these processes operate on fast time scales, others on slow ones. The fast processes, shown in green, relax rapidly into a quasi-equilibrium state, allowing their time scales to be eliminated from the description. Ultimately, the process operating on the slowest time scale, shown in red, governs the overall system dynamics. Hermann Haken termed this the \emph{enslaving principle} \cite[]{Haken-1983-synergetics}.

\cref{fig:limit}(c) presents an example from ecology. A seasonal drought, for instance, may stress the habitat of a small lake, yet once precipitation returns, the habitat recovers. This becomes critical when the recovery to the normal state becomes progressively slower \cite[]{Dakos-Carpenter-ea-2015-resilience}. Such critical slowing down serves as a measure of resilience in ecological systems: If the recovery time becomes sufficiently long, the system may lack the resilience to withstand a subsequent perturbation before full recovery has been achieved.

\section{The Blind Spot: Situation Awareness}
\label{sec:before-breakdown}

Before focusing on modeling, it is necessary to first address what \emph{cannot} be captured by formal models, namely the human element. System breakdown is partly caused by our inability to assess critical situations accurately. These psychological failure modes are far from random; they follow their own logic, of which we should be aware.

\textbf{Illusion of Control.}
Since the work of Langer (1975) \citep{Langer-1975}, it has been established that humans systematically overestimate their ability to influence critical situations. It is difficult to accept that certain external causes or random occurrences lie beyond our control, but it is even more difficult to distinguish what can be changed from what cannot. The ongoing debate surrounding climate change serves as a pertinent example.

\begin{figure}[t]
  \centering
  \includegraphics[width=0.45\textwidth]{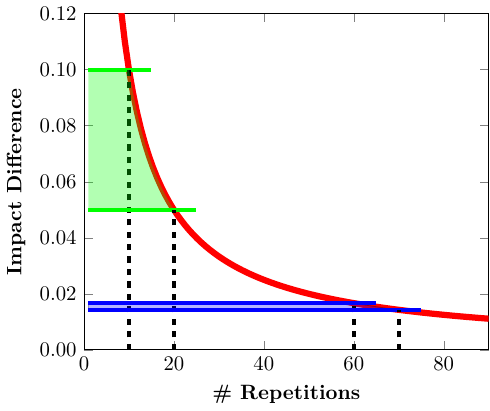}(a)
  \hfill
  \raisebox{25pt}{\includegraphics[width=0.45\textwidth]{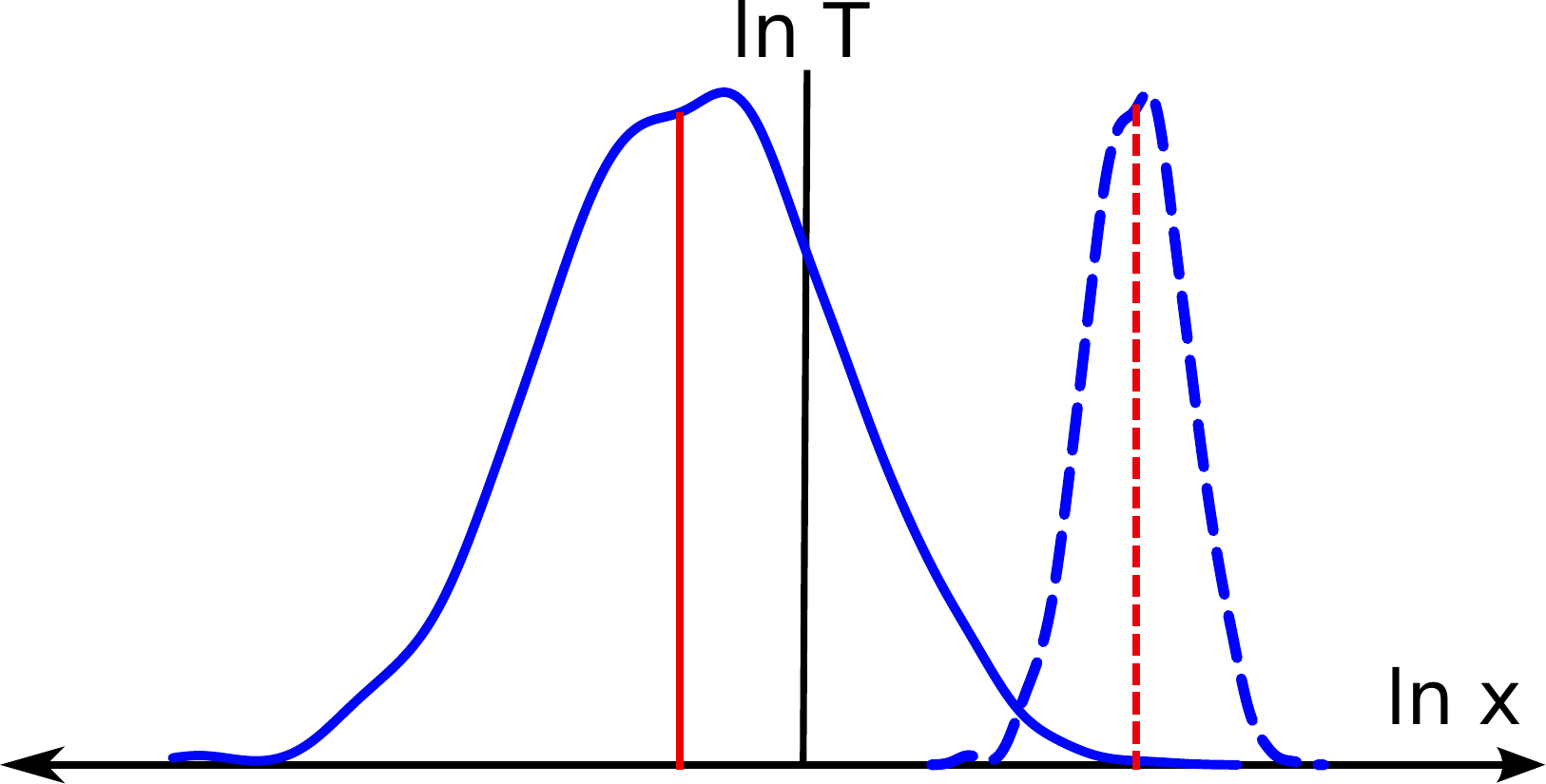}}(b)
  \caption{(a) Impact difference of repetitive information, following the Weber-Fechner Law, (b) Distribution of opinions $x$ (log-scale). $\ln T$ denotes the true answer. Solid lines sketch the distribution without social influence, dashed lines with  existing social influence \citep{Mavrodiev-Schweitzer-2021-enhanced,Mavrodiev-Schweitzer-2021}. 
  }
  \label{fig:cliff}
\end{figure}

\textbf{Poverty of Attention.}
Herbert Simon observed that a wealth of information  creates a poverty of attention \cite[]{Simon-1986-roleattentioncognition}. This phenomenon is readily experienced in the era of social media \citep{Huberman-Wu-2008}. However, it is not merely information overload that drives ignorance, but the overwhelming prevalence of repetitive negative messaging. The Weber--Fechner Law from psychophysics (see \cref{fig:cliff}a) offers an explanation for why the impact of such signals diminishes with increasing frequency of exposure. Alarmism is therefore counterproductive. Repeated invocations of imminent catastrophe rapidly lose their effect.

\textbf{Social Herding.}
Since the early observations of Galton (1905), numerous experiments have provided evidence for the so-called \enquote{Wisdom of Crowds} \cite[]{Surowiecki2005}. This is a genuine statistical effect that operates for well-defined questions with a determinate but unknown answer, such as \enquote{What is the length (in km) of the border between Switzerland and Italy?}\cite{Lorenz-Rauhut-2011-how-social}. The conditions require that individuals possess some prior knowledge of the subject and that they provide their estimates independently of one another. Under these conditions, the statistical mean over a large number of responses converges remarkably closely to the true answer.

The preconditions necessary for this effect to hold are, however, frequently disregarded. Two conditions are essential: a baseline level of individual competence, and the independence of responses. When individuals are exposed to the answers of others, they tend to incorporate this additional information into their own estimates, causing responses to converge over successive iterations \cite{Lorenz-Rauhut-2011-how-social}. As in opinion dynamics, consensus can emerge, yet the critical problem is that this consensus may converge on an answer that is objectively incorrect (see \cref{fig:cliff}b) \citep{Mavrodiev-Schweitzer-2021-enhanced,Mavrodiev-Schweitzer-2021}. Collective reinforcement thus leads to the shared adoption of a false conclusion, a well-known phenomenon that remains largely overlooked in public discourse.

\begin{figure}[t]
  \centering
\raisebox{30pt}{\fbox{\includegraphics[width=0.45\textwidth,trim=10mm 50mm 15mm 5mm, clip]{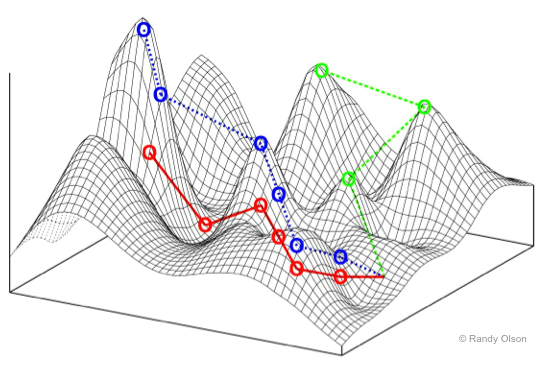}}}(a)
  \hfill
  \includegraphics[width=0.45\textwidth]{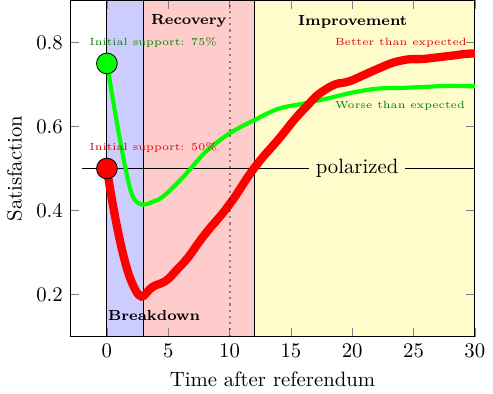}(b)
  \caption{(a) Evolutionary landscape: Higher peaks indicate better fitness (from Wikipedia). (b) Sketched dynamics of the population satisfaction after a polarized decision}
  \label{fig:brexit}
\end{figure}

\section{Wrong Expectations}
\label{sec:wrong}

The cumulative result of these psychological effects, among others, is a systematically distorted collective expectation regarding future developments.

Consider the evolutionary fitness landscape depicted in \cref{fig:brexit}(a). It resembles a mountain range, exhibiting several peaks of varying heights separated by deep valleys. The higher the peak, the greater the performance from a fitness perspective.

To move from a peak of moderate height to a superior one, it is immediately apparent that one must traverse a valley, and the higher the peaks, the deeper the intervening valleys. In other words, conditions must necessarily deteriorate before any improvement can be achieved.

The question is whether we accept that the same dynamics is applied to societal challenges. Consider the example of a country deciding on its future relationship with the European Union, a highly complex issue capable of polarizing an entire society. A public referendum is held, yielding a 50/50 result: 50 percent vote for closer integration with the EU, and 50 percent vote against it, with a narrow majority prevailing on the side of rejection.

The red curve in \cref{fig:brexit}(b) illustrates a schematic representation of the subsequent development of population-wide satisfaction over time. This is not empirical data, but a conceptual illustration. Following the referendum, a sharp decline in satisfaction is observed within a short period. This decline is attributable to the fact that 50 percent of the population was immediately dissatisfied with the collective outcome, having voted for closer EU relations. The remaining 50 percent, while aligned with the majority decision, had anticipated tangible improvements that did not materialize.

The country in question is Switzerland in 1992. The referendum concerned closer integration with the EU, and 50.3 percent of voters rejected it. The sketched satisfaction curve follows precisely the pattern described by the shock-breakdown-recovery framework introduced earlier. The shock in this case is polarization, a particularly severe shock, given its capacity to divide the entire population into opposing factions. Swiss society was required to absorb and recover from this disruption. The timescales involved are instructive: it was only after two years that the Swiss government and the EU started negotiations on their future relationship. Those negotiations, producing the so-called bilateral agreements, required six years to conclude. A further two years elapsed before new measures, such as the free movement of persons, came into effect in 2002. And it took approximately another decade before these measures translated into improvements perceptible to the Swiss population.

Now, more than 30 years later, a new referendum is being held on revised bilateral agreements with the EU. The prevailing sentiment is one of broad satisfaction with the existing arrangements. Hence, satisfaction with what has now become the status quo has increased substantially over this 30-year period.

This example illustrates the systematic error in expectations regarding short-term improvement. There is no direct path to the next peak; a deep valley must be crossed first. 
Second, the example supports the proposed framework of shock and recovery, demonstrating that severe polarization can function as the initiating shock, one that is not externally imposed, but self-generated through internal societal dynamics. Third, it provides an empirical reference point for the timescales associated with recovery and eventual improvement.

\section{Breakdown}
\label{sec:breakdown}

The general picture outlined here involves two prominent phases: breakdown and recovery.

The breakdown represents the dissipation phase, in which order is destroyed and disorder prevails. Established structures collapse, stock market value is destroyed, social security deteriorates into poverty, employment declines due to failing industries, and so forth. The prevailing signals are uniformly negative.

It offers little consolation to note that this process is an inherent feature of every evolutionary dynamic and therefore a component of future development. The economist Joseph Schumpeter coined the term \enquote{creative destruction.} All social and economic systems exist in a state of non-equilibrium rather than equilibrium, meaning they continuously test their own stability. When their current state becomes unstable, they transition into a new state that is more stable than the preceding one.

\begin{figure}[t]

  \centering
  
\raisebox{10pt}{\includegraphics[width=0.2\textwidth]{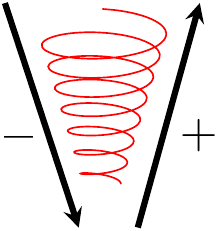}}(a)
\hfill
 \raisebox{10pt}{\includegraphics[width=0.38\textwidth,trim=30mm 0mm 0mm 0mm, clip]{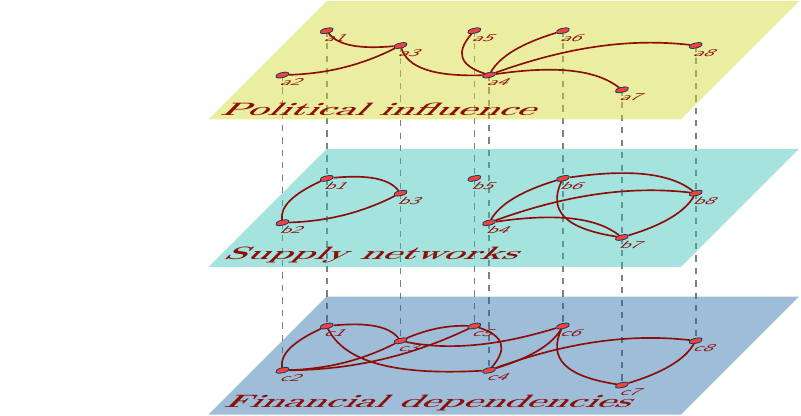}}(b)
  \hfill
  \includegraphics[width=0.28\textwidth]{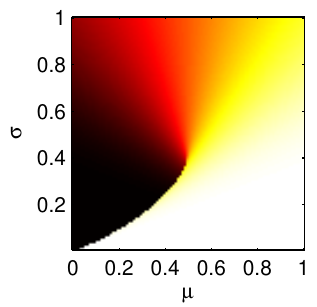}(c)
  \caption{(a) Positive feedback always amplifies the trend. (b) Multi-layer coupling of agents. (c) Phase diagram to demonstrate the impact of heterogeneity $\sigma$ and control parameter $\mu$ on the fraction of failed agents: The darker the color, the larger the fraction of failed agents. \citep{Lorenz-Battiston-Schweitzer-2009-systemic}.
    }
  \label{fig:heterog}
\end{figure}

Breakdown is typically accelerated by positive feedback loops, which amplify the prevailing trend regardless of its direction (see~\cref{fig:heterog}a).
When the trend is downward, positive feedback intensifies the decline. The National Health System during the COVID-19 crisis illustrates this clearly: a continuously shrinking nursing workforce was required to care for a continuously growing number of infected patients, a dual acceleration. The same acceleration dynamic was observed during the financial crisis of 2008.

The situation is further complicated by the fact that actors are coupled not only within a single layer but across multiple layers simultaneously (see~\cref{fig:heterog}b). Even when the dynamics within, for instance, the financial layer can be controlled, breakdown can propagate rapidly across the global supply layer or the political layer. A instructive example is the devastating tsunami that struck the east coast of Japan in 2011. It disabled three reactor blocks at the Fukushima nuclear power plant, but it also led to the immediate shutdown of seven nuclear power plants in Germany within a matter of days, followed by ten additional closures over the subsequent years.

This illustrates that while models exist for various types of failure cascades, a comprehensive description of how different interaction layers are coupled remains lacking, particularly when the political layer is involved.

Two principal mechanisms can mitigate such failure cascades. The first is heterogeneity, that is, the degree of difference among agents. Consider the diagram in \cref{fig:heterog}(c). The $x$-axis represents global stability, where zero denotes complete instability and one denotes complete stability. The $y$-axis represents agent heterogeneity, specifically how agents differ with respect to their susceptibility to a shock, analogous, for example, to individual resistance against a virus. The color scale indicates the fraction of the system that collapsed following a shock, such as the extent of viral spread, where white represents zero and black represents total system failure. A sharp boundary between white and black is clearly visible. This indicates that under certain conditions the entire system remains unaffected, yet a marginal change results in complete infection, a consequence of agents being so similar that they fail simultaneously. As heterogeneity increases, the color shifts to red and yellow, indicating that only a smaller fraction of the system fails. The reason is that sufficiently resistant agents withstand the shock, remain uninfected, and do not transmit the virus further \citep{Lorenz-Battiston-Schweitzer-2009-systemic}.

The same principle applies to banking networks. Bankruptcy cascades were halted precisely because certain banks held sufficient capital to remain solvent and therefore did not propagate losses to other institutions. This framework also clarifies the role of central banks: to identify which institutions require support in order to arrest the spread of financial contagion.

The second mitigation mechanism is decoupling, which is particularly relevant for supply networks. Globalization generates substantial economic benefits, but simultaneously propagates every production disruption across the entire supply network. Where it is feasible to reduce dependence on vulnerable supply sources and increase linkages to alternative ones, failure cascades can be contained \citep{Burkholz-Schweitzer-2019-international,Amico-Vaccario-Schweitzer-2024-efficiency,Amico-Verginer-ea-2024-adapting}.

\section{Recovery}
\label{sec:recovery}

Recovery always requires additional resources. A failed state cannot be transformed into a thriving democracy without cost. Where resources cannot be invested, improvement cannot be expected. However, support from system dynamics is available.

Perhaps the most surprising insight is that the same positive feedback processes that accelerated breakdown can be harnessed to accelerate recovery. Positive feedback amplifies the prevailing trend, regardless of its direction (see~\cref{fig:heterog}a). If a positive trend toward recovery can be induced, that trend becomes subject to amplification.

This implies that the focus should be on what system dynamics refers to as the \enquote{early symmetry break}. Initially, a symmetry exists between two possible directions: upward or downward. If the system can be pushed toward the upward branch, a positive development may follow.

The difficulty is that action must be taken \emph{early}, at a point when solutions are not yet clear and proposals are unlikely to command trust. In this situation, social feedback mechanisms can be leveraged. Two are of particular relevance.

\textbf{Reputation.} A demonstrated track record of guiding systems back onto a recovery path constitutes a powerful intangible asset — one capable of breaking the symmetry toward a positive development, because it provides a basis for trust \citep{Schweitzer-Mavrodiev-ea-2020-modelinguser,Schweitzer-2020-law-propor}.

\textbf{Reciprocity.} It is commonly understood as conditional cooperation: one cooperates only if the other party cooperates. Under this interpretation, if we  wait for the other party to move first, we loose the opportunity to shape the initial trend. It is considerably more effective to demonstrate cooperation unilaterally and use that signal to influence the other party's behavior \citep{Molm-2010,Falk-Fischbacher-2006}.

Why is failure so prevalent and successful recovery so rare?
First and foremost, the failed system, while performing poorly, is relatively stable (see \cref{fig:recover}b).
Consequently, recovery is not only costly but also slow.
Another reason lies in misconceptions about the nature of viable solutions. Following {Paul Watzlawick} \citep{Watzlawick-Weakland-Fisch-1974-change}, a distinction can be drawn between \emph{first-order} and \emph{second-order} solutions. A first-order solution is straightforward: more of the same. The German government may well represent the most consistent application of this approach. When industrial output declines, the response is to allocate subsidies. When output continues to decline, subsidies are increased. This strategy operates within narrow limits and remains viable only as long as the associated expenditure can be sustained.

A second-order solution is considerably more ambitious: it entails changing the system itself. The Argentine government provides a prominent example. {Federico Sturzenegger}, Argentina's Minister of Deregulation and State Transformation, similarly identifies inefficiencies in particular industrial sectors. However, in the absence of funds for subsidies, his approach is to eliminate all regulations in selected industries entirely, and subsequently to determine which regulations are genuinely necessary.

In both cases, the system is in an inferior state characterized by declining economic activity that is, unfortunately, highly stable and resistant to change. The difference lies in the objective: one approach seeks to restore a past \emph{status quo}, while the other seeks to increase the adaptivity of the system in the hope that it will settle into a new stable state (see also \cref{fig:quarter}c).

It cannot be determined in advance which strategy will prove more effective. Second-order solutions are, without question, considerably more risky. Governments and citizens alike tend to prefer the familiar stability of the past over open-ended evolution.

\section{The New Normal: No Time Scale Separation}
\label{sec:vuca}

\begin{figure}[t]
  \centering
    \includegraphics[width=0.44\textwidth]{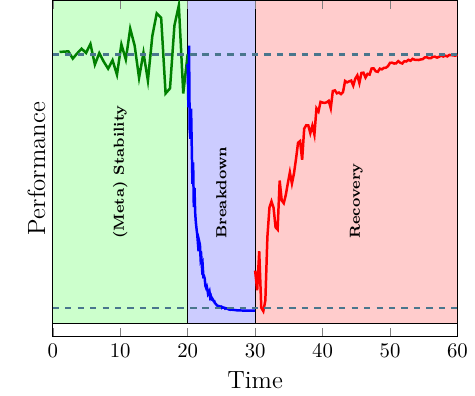}(a)
\hfill
\includegraphics[width=0.44\textwidth]{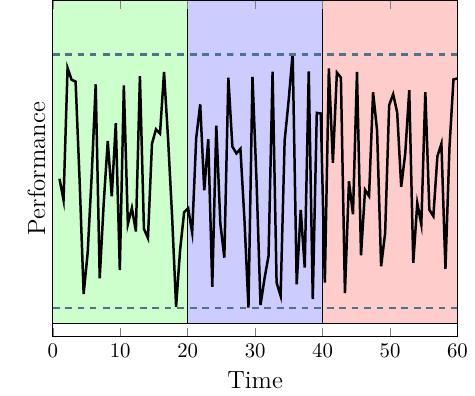}(c)
\bigskip

\hspace*{0.7cm}    \fbox{\includegraphics[width=0.35\textwidth]{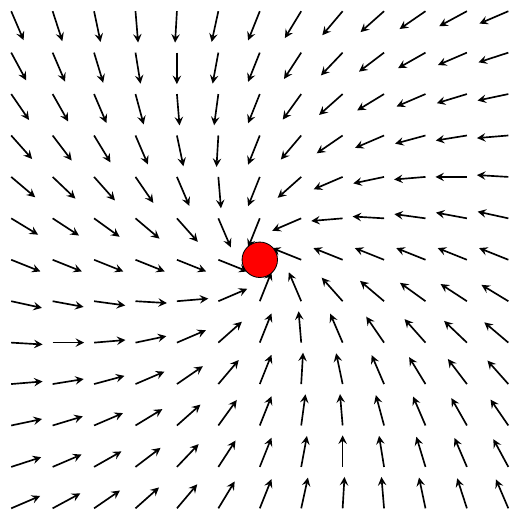}}(b)
\hfill
\fbox{\includegraphics[width=0.35\textwidth,angle=90]{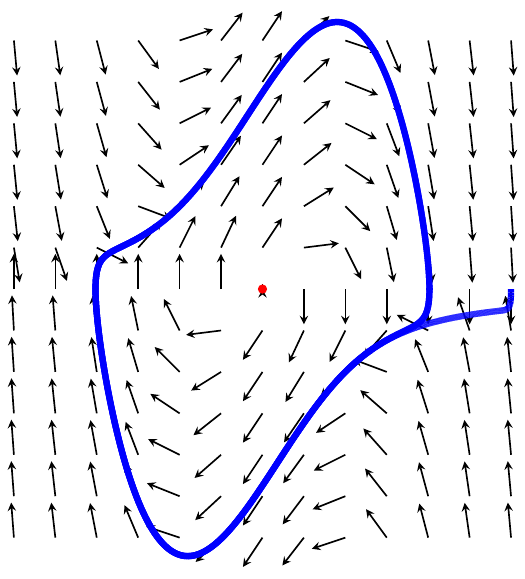}}(d)
\caption{(a,b) If a normal state can be clearly identified and time scales for breakdown and recovery can be separated, the system is resilient if its dynamics approaches a fixed point. (c,d) In volatile systems without time scale separation, the system is resilient if its dynamics follows a life cycle, instead.}
  \label{fig:normal}
\end{figure}

Having established an understanding of the three distinct phases shown in \cref{fig:normal}(a), we now turn to the more realistic scenario presented in \cref{fig:normal}(c).
Here, no distinct phases can be identified and no separation of time scales is observed.
Instead, the system exhibits highly fluctuating performance that never settles into a stationary state.
This is the typical behavior of most organizations, described in management science as the VUCA state \cite[]{Mack-Khare-2016-perspectivesvucaworld}:
V stands for \emph{Volatility}, referring to rapid, stochastic changes.
U stands for \emph{Uncertainty}, referring to the unpredictability of events.
C stands for \emph{Complexity}, encompassing many agents with interdependent relationships and emergent system properties.
A stands for \emph{Ambiguity}, meaning that neither the state of the system nor the appropriate corrective measures can be unambiguously interpreted.

This dynamics is characteristic of social organizations, in which a \enquote{normal} reference state cannot be defined.
Instead, the system resides in a state of permanent \emph{non-equilibrium}.
Large fluctuations in performance arise because various processes operate concurrently on the same time scale: entry and exit dynamics of collaborators, changes in the division of responsibilities, temporal variations in task assignments, fluctuations in the acquisition of funded projects, and so forth.
As a result, \enquote{breakdowns} of varying magnitudes occur frequently and are followed by rapid recoveries, with the next \enquote{shock} typically occurring before the system has fully recovered from the previous one.
Volatility, in this sense, is the new normal.

What are the implications of this observation for the concept of resilience?
Comparing \cref{fig:normal}(a-c) with the vector field plots in \cref{fig:normal}(b-d) is instructive.
In \cref{fig:normal}(a), where three separated time scales are present, resilience implies that the system returns to a relatively stable state following breakdown and recovery.
In dynamical systems theory, this corresponds to a focal point or fixed point, as illustrated in \cref{fig:normal}(b), reflecting stability and functionality under the assumption that target values are attainable.

The highly volatile system depicted in \cref{fig:normal}(c) possesses no such fixed point to return to.
This does not, however, imply an absence of resilience.
Other stable solutions exist, most notably the limit cycle shown in \cref{fig:normal}(d).
In this regime, the system evolves continuously along a closed trajectory: performance, stability, and adaptivity all vary considerably.
Despite persistent fluctuations, the system remains functional, it is \emph{still there}.
In this case, resilience is expressed through \emph{life cycles} rather than fixed points.
If the system is subjected to external shocks, it will return to the limit cycle shown in \cref{fig:normal}(d) in the same manner as it returns to the fixed point shown in \cref{fig:normal}(b).

The key conclusion is that a quantitative framework for resilience must account for these two distinct scenarios.
When time scales can be separated, macro-level modeling using system dynamics methods is feasible, as descriptive macro variables can be defined to characterize target values.
When time scales overlap, breakdown and recovery processes unfold simultaneously and are mutually intertwined.
Then, agent-based models are required to elucidate how robustness and adaptivity interact to produce resilience as an \emph{emergent property} of the system.

\section{Foundation of Resilience}
\label{sec:resilience}

With the VUCA framework in mind, it becomes evident that breakdown and recovery are two aspects of a single phenomenon, best captured by the term \emph{Change}. Resilience can therefore be defined as the capacity to {maintain} functionality when facing change.
The two dimensions of resilience are still  foundational: the \emph{structural} dimension, which corresponds to \emph{robustness}, and the \emph{dynamic} dimension, which corresponds to \emph{adaptivity}.

\begin{figure}[t]
      \centering
  \begin{minipage}[c][][c]{0.45\linewidth}
    \includegraphics[width=0.99\textwidth]{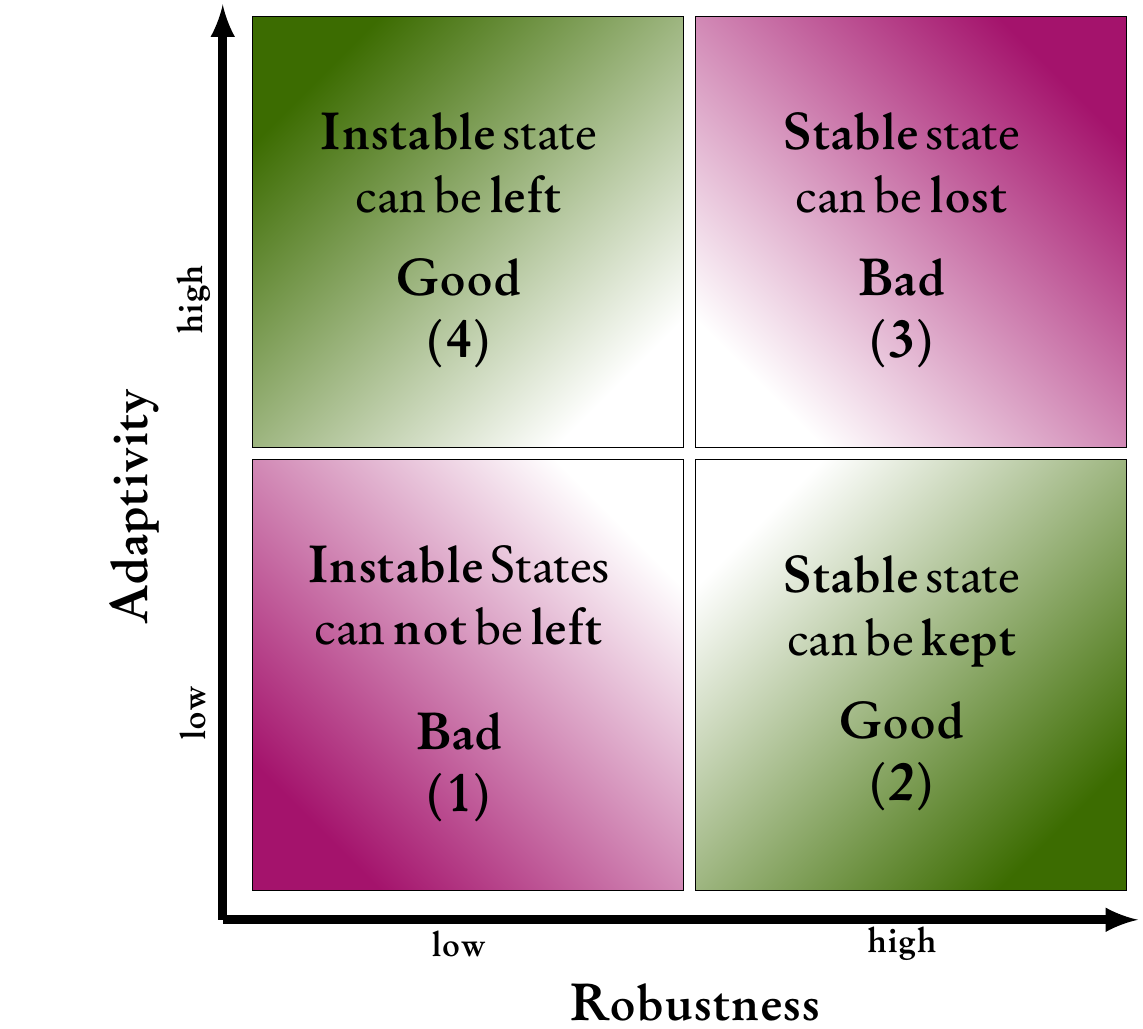}(a)
  \end{minipage}
  \hspace*{2cm}
  \begin{minipage}[c][][c]{0.25\linewidth}
       \includegraphics[width=0.99\textwidth]{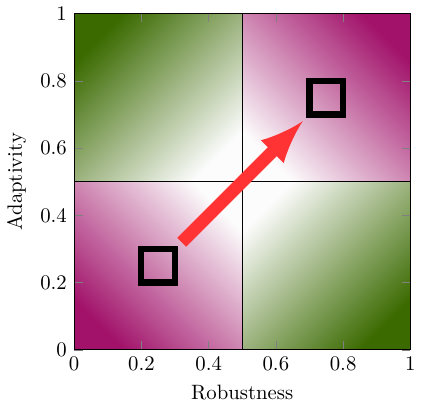}(b)
       \medskip
       
    \includegraphics[width=0.99\textwidth]{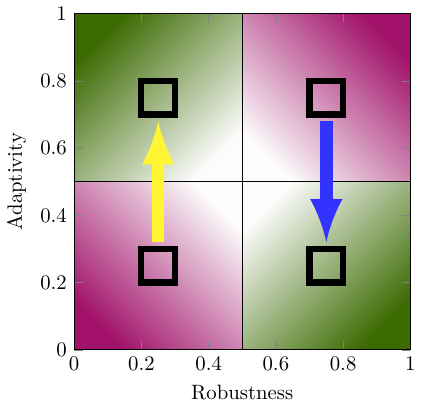}(c)
  \end{minipage}
  \caption{Resilience (green: high, red: low) as a function of robustness and adaptivity. (a) Good and bad states. (b) 1st order solution. (c) 2nd order solution \citep{Schweitzer-Andres-ea-2022-modelingsocialresilience,Schweitzer-Zingg-Casiraghi-2023-strugglingwithchange}.}
  \label{fig:quarter}
\end{figure}

The relationship between these two dimensions and overall resilience is illustrated in \cref{fig:quarter}(a), where green indicates high resilience and red indicates low resilience.
Beginning in the lower-left region of the graph: a system exhibiting both low robustness and low adaptivity also exhibits low resilience. This follows from the fact that such a system lacks the structural integrity to absorb shocks while simultaneously lacking the adaptive capacity to recover from an inefficient state. Consequently, a VUCA system with low robustness must compensate through sufficient adaptivity to cope with change.

In the upper-right region, a system combines high robustness with high adaptivity, yet this configuration is not without risk. The pressure to exploit high adaptive capacity can lead management to implement continuous organizational changes, revising management strategies, restructuring hierarchies, and redistributing responsibilities across divisions. This pattern of uncoordinated overactivity risks eroding the very robustness the system had previously achieved.

Change, therefore, is a double-edged sword: it is necessary to escape unfavorable states, but it must not be allowed to destabilize favorable ones.

A further implication of this framework is that first-order solutions are insufficient for improving resilience. Independently increasing both robustness and adaptivity
merely produces a system that scores highly on both dimensions, but it does not yield a more resilient system, as shown in \cref{fig:quarter}(b).

Achieving a genuinely more resilient state, shown in \cref{fig:quarter}(c), requires second-order solutions because
resilience is reducible neither to stability nor to flexibility alone. Stability enables a system to withstand shocks but does not facilitate recovery from them. Flexibility may allow rapid restructuring, but the more a system adapts to a specific situation, the less prepared it becomes to absorb shocks arriving from unforeseen directions. Resilience, properly understood, denotes the capacity to remain prepared for any perturbation while retaining the ability to recover.

The counterintuitive interplay between adaptivity and robustness has a deeper structural explanation: every intervention produces intended \emph{and} unintended consequences. Intended consequences drive the system toward the desired state and are therefore stabilizing; unintended consequences drive it away from the desired state and are therefore destabilizing. Resilience represents a deliberate compromise, an orientation that does not focus exclusively on achieving a desired state, but that also accounts for the destabilizing potential of unintended consequences.

\section{Generative Micro-Models of Resilience}
\label{sec:gener-micro-models}

Modeling resilience for VUCA systems requires detailed knowledge of the system at the micro level. Unlike ecological systems, for instance, there is no definitive macro-level variable that captures the system state, such as biomass production. Instead, a generative model is needed in which resilience emerges as a property of interacting agents. This necessitates information about both the agents and their relations, requiring fine-grained data.

\begin{figure}[t]
  \centering
  \includegraphics[width=0.99\textwidth]{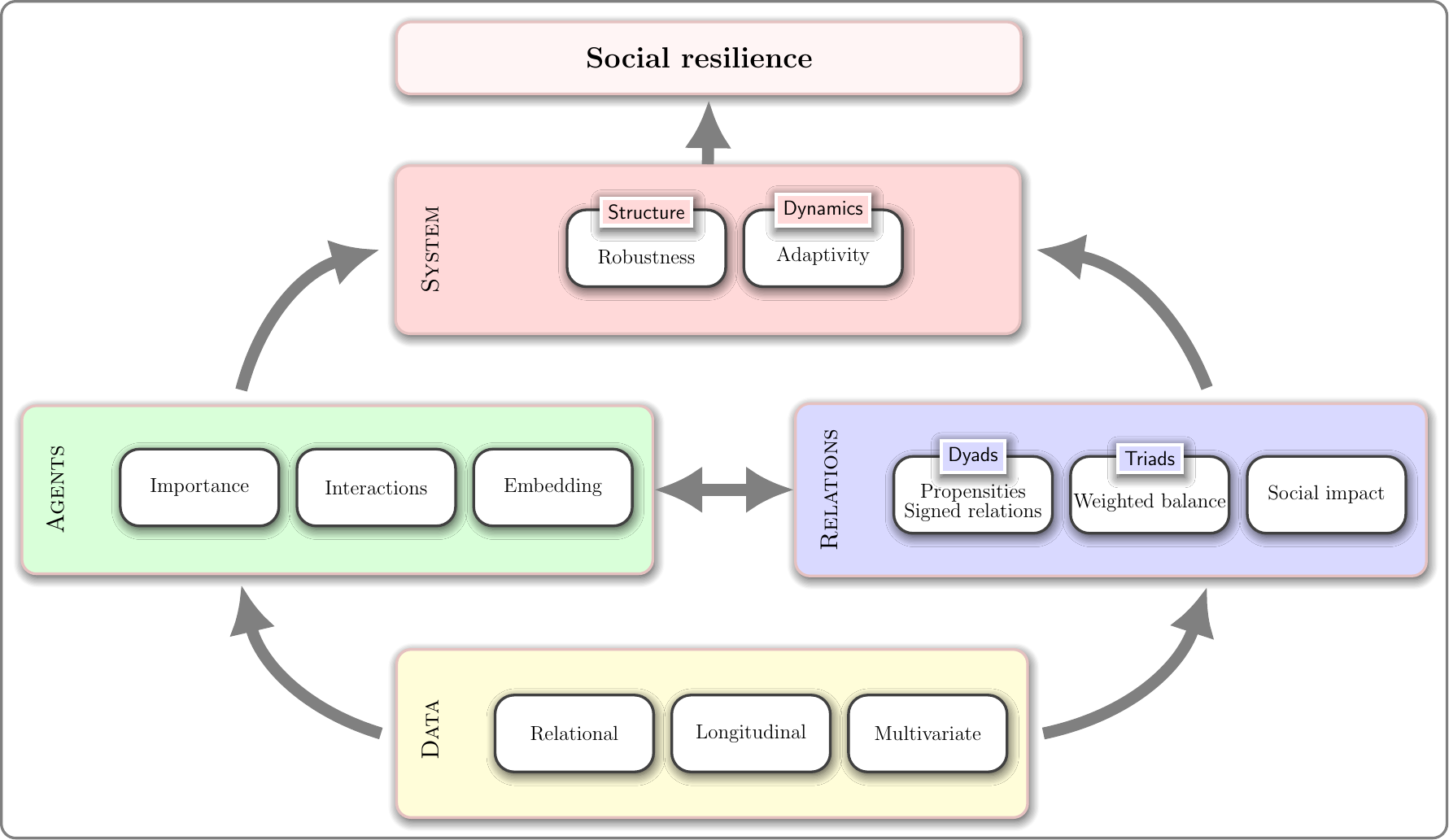}
  \caption{Operationalization to calculate social resilience bottom up (from \citep{Schweitzer-Andres-ea-2022-modelingsocialresilience})}
  \label{fig:overview}
\end{figure}

A comprehensive treatment of such micro-level models is discussed in \citep{Schweitzer-Andres-ea-2022-modelingsocialresilience}.
The general framework is illustrated in \cref{fig:overview}. Following a bottom-up approach, the framework begins at the data layer with heterogeneous data sources. This data-driven approach enables the specification of both agents and their interactions, which are subsequently integrated into a system model. Within this model, the concepts of robustness and adaptivity can be formally defined for the system under consideration, for the details see \citep{Helfgott2018,Schweitzer-Andres-ea-2022-modelingsocialresilience,Schweitzer-Zingg-Casiraghi-2023-strugglingwithchange}. Based on the quantification of robustness and adaptivity, resilience can ultimately be derived for a given social organization.

The following sections provide examples of how generative models can be informed by various data sources and the respective tools available for extracting information from them.
Collaboration networks were selected as the primary case study because analogous examples can be found across numerous domains: scientists collaborate to produce publications, developers collaborate to produce open-source software, firms collaborate to produce patents, and political legislators collaborate within parliamentary bodies to produce bills and laws.

Collaboration networks are susceptible to failure, and their breakdown is frequently self-inflicted. Contributions to the common good are not uniformly distributed among participants; free riders who benefit from the efforts of others without contributing equivalently are a common phenomenon. Furthermore, agents exhibit heterogeneous motivations for contributing to collaborative endeavors.

Data-driven modeling of such systems requires data at two distinct levels: the collective output, that is, the product itself, must be characterized, as must the agents responsible for generating that output through their collaboration. The objective is not only to analyze the resilience of these collaboration networks, but also to demonstrate how agents and their interrelations can be systematically quantified.

    \section{Inform Micro-Models}
\label{sec:inform-micro-models}

\begin{figure}[t]
  \centering
  \includegraphics[width=0.5\textwidth]{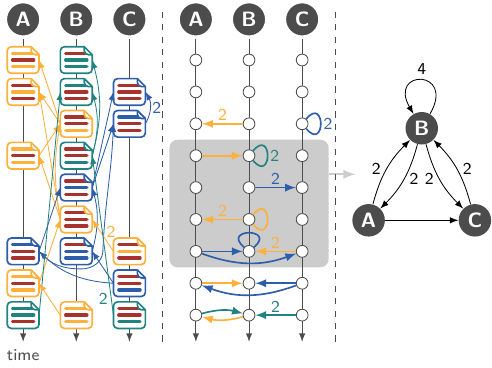}(a)
  \hfill
  \includegraphics[width=0.4\textwidth]{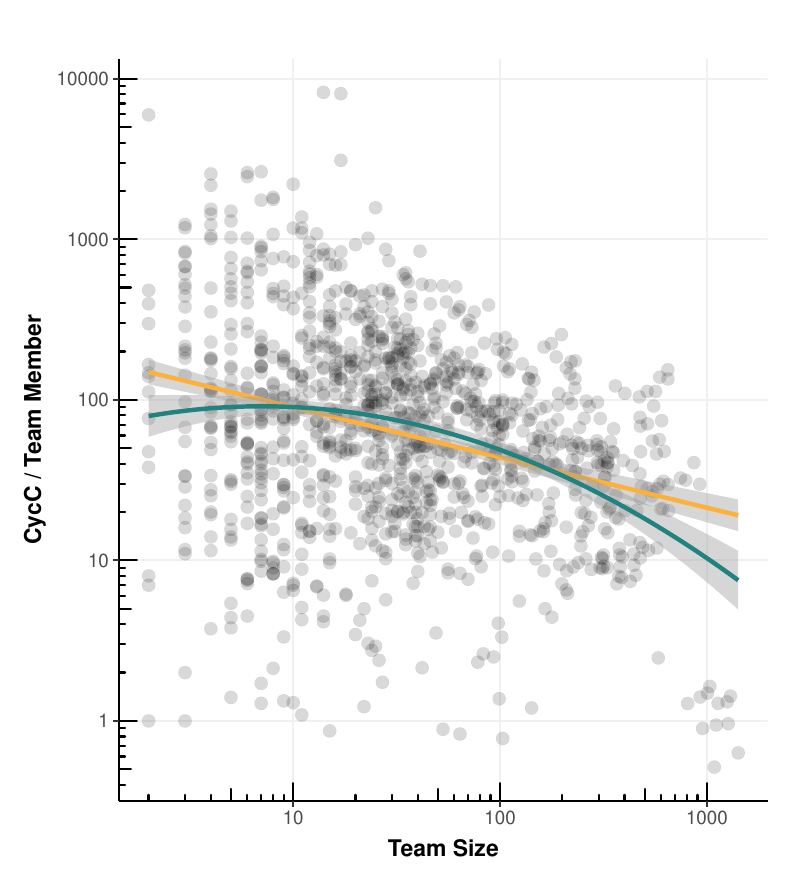}(b)
  \caption{(a) Git2net identifies which developer wrote which code, who revised or improved it, and the relative contribution of each developer to the project.\citep{Gote-Scholtes-Schweitzer-2021-analysingtime,Gote-Scholtes-Schweitzer-2019-opensource} (b) Productivity decline vs. team size. The contribution per team member is quantified by cyclomatic complexity, a standard measure of software complexity. The log-log plot displays two curves, corresponding to a linear and a quadratic model, respectively. From the quadratic model an optimal team size between 7 and 19 developers is identified. To date, this represents the largest study conducted on this topic. \citep{Gote-Mavrodiev-ea-2022-big}
}
  \label{fig:oss}
\end{figure}

\textbf{Mining GIT Repositories.}
Open source software projects were selected precisely because their git repositories provide the necessary information about both the software and the developers. A software tool, Git2net, was developed for the process of data extraction, as sketched in \cref{fig:oss}(a) \citep{Gote-Scholtes-Schweitzer-2021-analysingtime,Gote-Scholtes-Schweitzer-2019-opensource} (see also \cite[]{Gote2021}). 
From these data, the social network of collaborations can be constructed as a time-stamped interaction network, rather than a simple aggregated network. The dynamics of the software results in a directed acyclic graph (DAG), in which each path represents a sequence of consecutive co-editing relationships among developers working on a given file.

Of particular interest was the dependence of productivity per team member on team size. %
The results are shown in \cref{fig:oss}(b). A clear negative relationship between team size and productivity per member is observed, consistent with the Ringelmann effect in social psychology and Brooks' law in software engineering \citep{Gote-Mavrodiev-ea-2022-big}. This decline in productivity is attributed to coordination overhead that scales with team size, reflecting the increasing complexity of collaboration within larger teams. Consequently, coordination overhead and the measures taken to mitigate it are likely to have implications for the resilience of the project, as discussed in \cref{sec:discussion}.

\begin{figure}[t]
  \centering
    \includegraphics[width=0.45\textwidth]{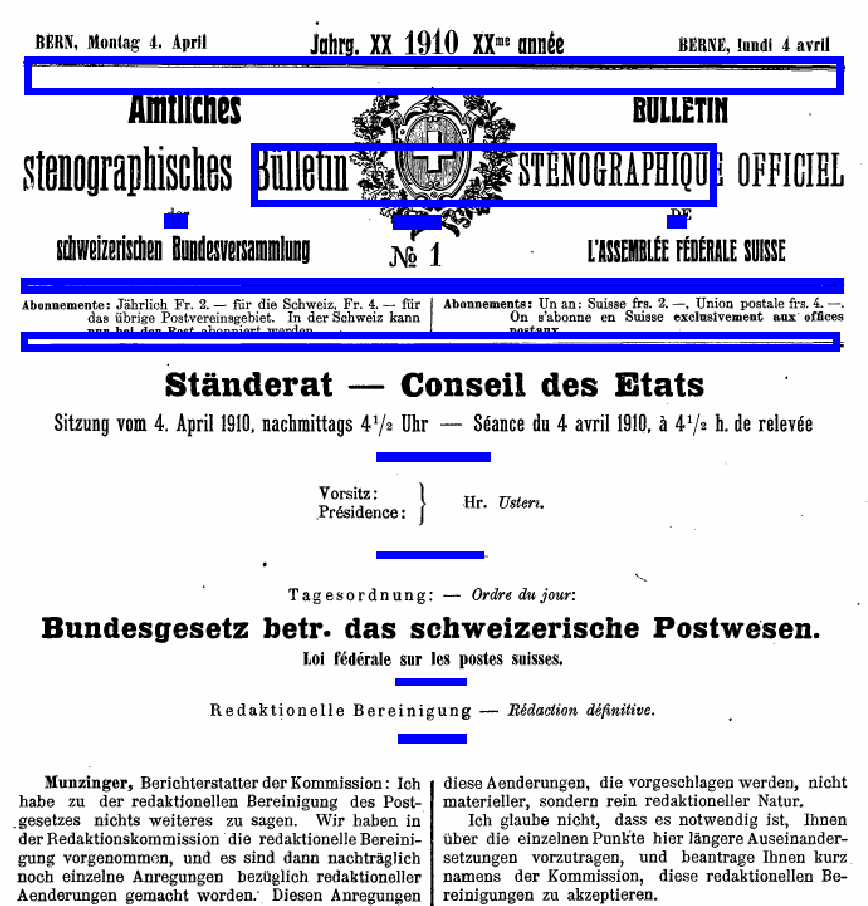}(a)   
  \hfill
  \includegraphics[width=0.45\textwidth]{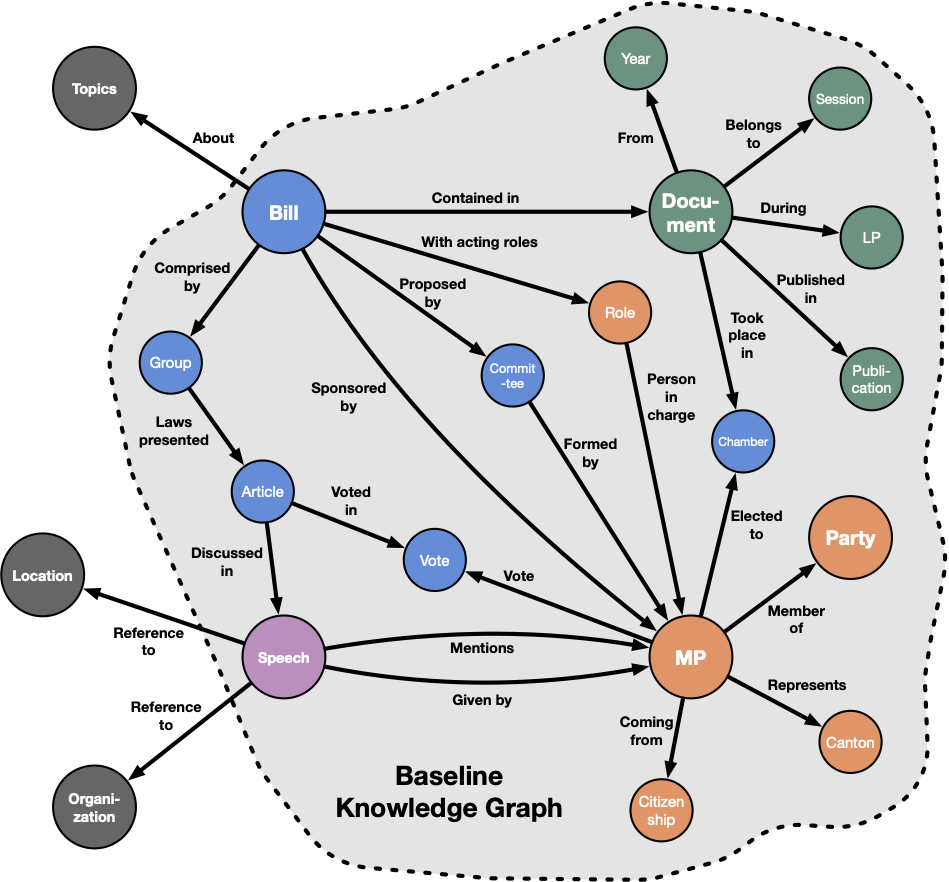}(b)
  \caption{(a) Complex layout of an official document from the Swiss parliament \citep{Salamanca-Brandenberger-ea-2024-processinglarge}. (b) Basic structure of knowledge graph with different entities, relations and attributes. 
  }
  \label{fig:KG}
\end{figure}

\textbf{Knowledge Graphs.}
Data from OSS projects primarily addresses syntactic aspects, structure and relations between elements, rather than semantic aspects, i.e., meaning and interpretation. When analyzing social or political organizations, a more elaborate representation of the semantic layer is required.

One approach to addressing this challenge is the knowledge graph, a graph-based database that represents different types of entities and their relations. As an example, we consider the Swiss parliamentary proceedings, for which a complete record of documents spanning 130 years is available \citep{Salamanca-Brandenberger-ea-2024-processinglarge}.

These documents consist of printed material, frequently exhibiting a complex layout structure as \cref{fig:KG}(a) shows. Consequently, data extraction requires considerable effort, particularly since all extracted content must be manually corrected and proofread. The resulting knowledge graph contains all members of parliament along with their speeches, interventions, bills, and associated metadata. The edges of the graph represent relations such as contributions to a bill or membership in a committee. A schematic representation of such a network is shown in \cref{fig:KG}(b).

The construction of a knowledge graph generally comprises three major steps \citep{Salamanca-Brandenberger-ea-2024-processinglarge}. First, numerous data mining challenges must be addressed in order to identify the structure and content of more than 200,000 pages of parliamentary proceedings. Second, the extracted documents are processed using named entity recognition tools to obtain embeddings and perform topic modeling. Third, the resulting knowledge graph must be made accessible through a web-based interface and a REST API, enabling user-friendly manipulation and analysis of the data.

Given this extensive graph database, various projections can be employed for further analysis. For instance, one can examine how different topics are interconnected, or which political parties supported specific proposals, over a time horizon of 130 years. This makes it possible to analyze, for instance, the political position of Switzerland before and during the First and Second World Wars, at a level of granularity that was previously unattainable.

\begin{figure}[t]
  \centering
	\includegraphics[width=0.99\textwidth, height=8cm,keepaspectratio]{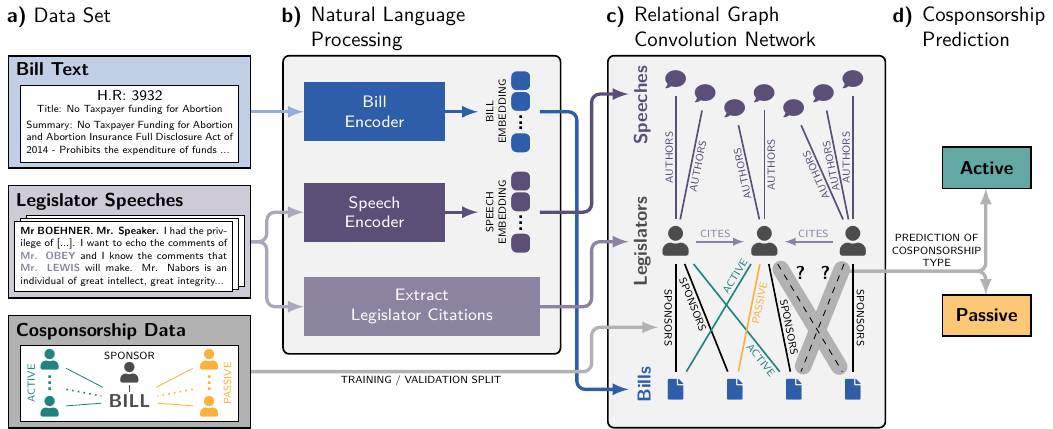}%
    \caption{Overview of the workbench for analyzing active and passive cosponsorship in the US Congress \citep{Russo-Gote-ea-2023-helpingfrienda}.}
  \label{fig:Congress}
\end{figure}

\textbf{Large Language Models.}
The knowledge graph alone is not sufficient to model agents' decisions. Taking parliamentary members as agents, for instance, additional information about their \enquote{ideology} would be required. This is a \emph{latent variable}, i.e., one that cannot be assessed directly. It must therefore be inferred from these members' contributions to bills, their speeches, their activity in committees, their party role, etc \citep{Russo-Gote-ea-2023-helpingfrienda}.

Several tools from artificial intelligence now allow for more complex analyses. These tools, commonly abbreviated as BERT, LLM, or GPT, essentially extend natural language processing (NLP), which was already employed for the knowledge graph. To illustrate their capabilities, the US Congress serves as an example \citep{Russo-Gote-ea-2023-helpingfrienda}. When examining the outcomes of roll-call votes, a strong polarization between Republicans and Democrats can be observed. In contrast, a remarkable degree of support and cooperation is evident during the drafting of bills. In the 115th US Congress, 15,000 bills were introduced within four years. Given this vast number, garnering support for a bill early on is essential for it to be considered. Such support cannot come solely from members of one's own party; involvement from the opposition is also required.

Two forms of cooperation can be observed: active cosponsorship, in which a legislator is already involved in the preparation phase and drafting of a bill, and passive cosponsorship, in which a legislator supports the bill only by signing it once completed. Both forms of support occur on a bipartisan basis.

\cref{fig:Congress}~ presents a schematic overview of our framework for representing legislators, which consists of four steps.
Step 1 involves data acquisition. Data on cosponsorship, bills, and speeches were collected over an eight-year period (112th to 115th Congress), comprising information from 50,000 bills and transcripts of 120,000 speeches.

Step 2 involves generating high-dimensional embeddings. For this purpose, separate encoders were developed for bills and speeches. In addition, citations, i.e., references indicating who cites whom, were detected and extracted from both bills and speeches.

In Step 3, a relational graph convolutional network was developed. This is a specialized type of graph neural network used to learn representations. It consists of three layers with distinct sets of nodes, speeches, legislators, and bills, and a set of edges representing authorship of speeches, citation of legislators, and sponsorship of bills (both active and passive). The result of Step 3 is what we term a holistic representation of legislators.

In Step 4, the model is trained and evaluated against baseline models, such as random forest classifiers and simple feedforward neural networks. This step serves to demonstrate that the proposed model outperforms these baseline approaches.

Finally, the model is used to predict cosponsorship. Our results \citep{Russo-Gote-ea-2023-helpingfrienda} indicate that in cases of active cosponsorship, the cosponsor primarily supports the sponsor, that is, the person, whereas in cases of passive cosponsorship, the cosponsor primarily supports the bill, that is, its content. This finding offers valuable insight into the underlying motivations of legislators when supporting bills.

\section{Resilience and Performance}
\label{sec:discussion}

Informed micro-models allow the study of robustness, adaptivity, and performance in volatile systems. Compared to, for instance, parliamentary systems, Open Source Software projects have the advantage that their dynamics are well documented in GIT repositories, and performance can be quantified through developer productivity. Consequently, the relationship between performance and resilience in volatile systems can be examined in considerable detail.

The underlying tension in this relationship is already suggested by the empirical observation that productivity per developer declines as teams grow. While the growth of the team size may be interpreted as an indication of successful development, the concomitant nonlinear increase in coordination effort constitutes an unintended consequence. This ultimately leads to conditions that threaten the viability of the project.

\begin{figure}[t]
  \centering
  \includegraphics[width=0.43\textwidth]{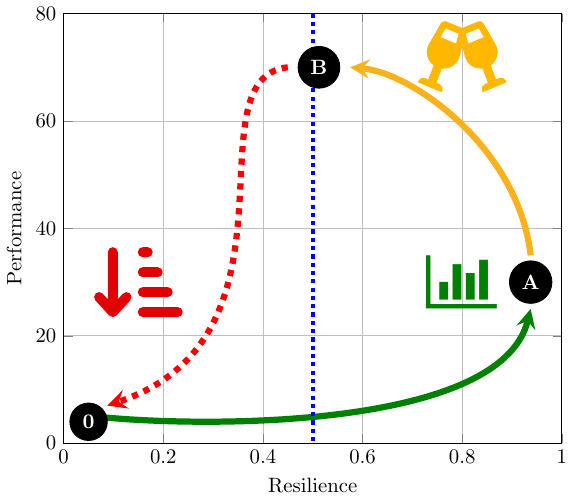}(a)
  \hfill
      \includegraphics[width=0.47\textwidth]{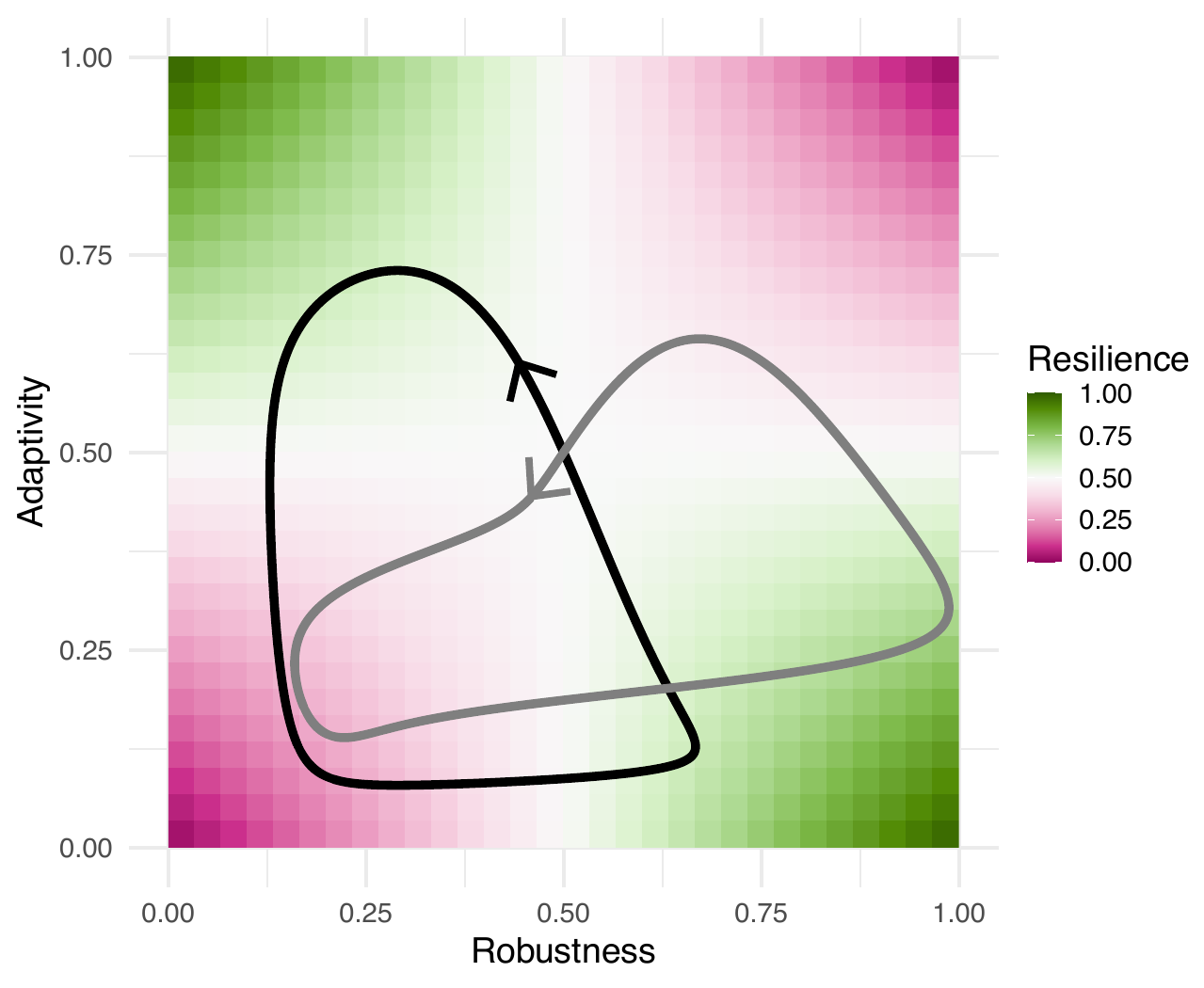}(b)
  \caption{
The Resilience Life Cycle: 
(a) The $x$-axis represents resilience, while the $y$-axis represents performance. (b) The $x$-axis represents robustness, the $y$-axis adaptivity and the color code high (green) and low (red) resilience \citep{Schweitzer-Zingg-Casiraghi-2023-strugglingwithchange}.}
  \label{fig:lifecycle}
\end{figure}

As an illustrative example, a specific open-source software project was analyzed over a 10-year period \citep{Casiraghi-Zingg-Schweitzer-2021-downsideheterogeneity,Garcia-Zanetti-Schweitzer-2013,Zanetti-Scholtes-ea-2013}.
Its resilience life cycle is shown in \cref{fig:lifecycle}(b), using the dimensions of robustness and adaptivity \citep{Schweitzer-Zingg-Casiraghi-2023-strugglingwithchange}.  
In its early stages, the collaboration network exhibited a relatively balanced structure
(see also the snapshots in Figure 17a-c in \citep{Schweitzer-Andres-ea-2022-modelingsocialresilience}). 
As the project expanded, the accompanying increase in coordination overhead necessitated the appointment of a project manager to address this growing complexity. The resulting collaboration network, however, displayed an unstable configuration dominated by a single highly central node, entailing a considerable risk: any disruption affecting this individual could jeopardize the entire project. This risk materialized when internal conflicts over responsibilities and hierarchical structures led to the removal of the central manager, representing a major shock and a serious threat to the project's continuity. Nevertheless, the project subsequently recovered, providing evidence of its resilience. The resulting collaboration network then reverted to a more balanced structure, reflecting a more equitable distribution of tasks.

Shifting focus from robustness to productivity during the period of instability, all performance measures increased while robustness declined. This observation motivated the hypothesis that a tradeoff exists between higher productivity and lower robustness, which was subsequently tested in a large-scale study comparing multiple measures of productivity and system resilience.

Prior to the breakdown, a critical slowing down in the system dynamics was observed, analogous to critical slowing down documented in ecological systems. This indicates that the system exhibited high productivity but low resilience. Following the shock, resilience increased and the system stabilized, albeit at the cost of diminished performance. Only during the recovery phase did a broader reorganization of the project occur: resilience temporarily declined again as the project progressed toward recovery, before eventually stabilizing at an acceptable level of performance, though not at its peak value. Furthermore, it was confirmed that network-based and productivity measures, beyond high productivity alone, can serve as early warning indicators of project failure.

These insights can be summarized in the mockup shown in \cref{fig:lifecycle}(a). 
At the outset of a new project, the system is located at point 0, characterized by low performance and low resilience. Robustness and adaptivity increase only once certain structures have been established, for example, a social network, and dynamic processes, such as collaborations, operate on this network. Point A represents an optimal balance between resilience and performance.

However, organizational pressure to improve performance often persists. As a result, resources originally allocated to maintaining resilience are frequently redirected toward further increasing performance. A well-known example is the German railway company Deutsche Bahn, which planned an initial public offering (IPO). To present improved performance indicators, safety margins and redundant fallback structures were reduced. Performance indeed increased, but resilience declined correspondingly. At point B, this apparent success was widely celebrated. Subsequently, however, the system collapsed, as it was no longer sufficiently robust.

Interviews with managers reveal that such strategies are considered rational from their perspective, since compensation structures reward performance rather than resilience. This asymmetry exists largely because adequate resilience measures for dynamic organizations have not yet been established. This gap has motivated much of our research, which focuses on the development of such resilience measures.

Notably, Deutsche Bahn continues to operate, illustrating that a resilience life cycle exists, allowing systems to recover and restart. Germany has since initiated a large-scale infrastructure program to rebuild resilience within its railway system, with estimates suggesting that returning to point A may take more than a decade.

        \section{Conclusions}
\label{sec:conclusions}

Resilience is more than a mere buzzword.
It combines two dimensions: the ability to withstand shocks and the ability to recover from them.
In this respect, resilience differs fundamentally from stability.

Perhaps the most difficult aspect to accept is that resilience necessarily entails a trade-off.
Resilient systems cannot maximize any single systemic feature, such as stability or performance, because they must simultaneously retain the capacity to cope with various shocks.
As Confucius observed:
\enquote{Our greatest glory is not in never failing, but in rising every time we fail.}

This requirement becomes particularly evident in so-called VUCA systems.
These volatile, highly complex systems do not permit a clear separation between periods of normalcy and periods of shock, breakdown, or recovery.
As a consequence, the nature of resilience shifts from being a state to being a life cycle.
Modeling such systems requires generative agent-based or network models, in which resilience constitutes an emergent property of the system.

This life-cycle dynamic arises from a number of causes.
Most importantly, social organizations quite often generate the causes of their own failure.
The so-called human factor plays a considerable role in this respect.
Thus, the systemic risk of breakdown is \emph{endogenous} to the system.
The \enquote{failure of the few} is amplified by various positive feedback processes, such that large cascades become possible.
This dynamic is often referred to as \emph{self-organized criticality}.
Evolving systems tend to favor such vulnerable states, as they often enhance performance.
Consequently, a tension between high resilience and high performance must be accepted.

Another important reason for the resilience life cycle is the systematic generation of unintended consequences.
It is difficult to foresee, and even more difficult to accept, that all stabilizing measures induce counteracting reactions that destabilize the system.
Evolving systems, including social and economic ones, exist in a state of permanent non-equilibrium.
This means that they require considerable resources to maintain their current state and continually test their own stability, thereby tending to move toward a more stable but less efficient state.

Accordingly, from a systemic perspective, the objective is not to maximize present performance but rather to prepare for future change.
To this end, thinking beyond established paradigms proves more valuable than attempting to reconstruct the past.

\section*{Acknowledgements}
\label{sec:acknowledgements}

I wish to thank my co-authors and my collaborators at the Chair of Systems Design of ETH Zurich.
Particular thanks goes to  {Laurence Brandenberger} and {Sophia Schlosser}, as well as to 
{Ingo Scholtes}.
I also want to thank 
      Rebekka Burkholz,       Giona Casirahi,        David Garcia,     Christoph Gote,       Pavlin Mavrodiev,  Giuseppe Russo,       Luis Salamanca, and Christian Zingg.  
Information about publications and projects can be found at \url{sg.ethz.ch}.

\end{document}